# Micro and Nano 3D investigation of complex interplay between gut alterations and dementia

F. Palermo[1], N. Marrocco[1], L. Dacomo[2], E. Grisafi[2], M. Musella[1], V. Moresi[1], A. Sanna[1], L. Massimi[1], I. Bukreeva[1], O. Junemann [1], I. Viola[1] M. Eckermann[3], P. Cloetens[3], T. Weitkamp[4], G. Gigli[1], G. Logroscino[5] N. Kerlero de Rosbo[1], C. Balducci[2*] and A. Cedola[1*]
*equal contribution

1 CNR-Nanotec Rome c/o Phys.Dip. SAPIENZA University
2 Istituto di Ricerche Farmacologiche Mario Negri IRCCS, Milan, Italy
3 Synchrotron ESRF, Grenoble France.
4 Synchrotron SOLEIL, Saint-Aubin, France.
5 University "Aldo Moro" Bari and Centro di Malattie Neurodegenerative e invecchiamento cerebrale & Department of clinical research in neurology c/o Fondazione Card. G. Panico di Tricase.

**Abstract**

Alzheimer's disease (AD), a debilitating neurodegenerative disorder, remains one of the foremost public health challenges of our time. Despite decades of research, its etiology largely remains enigmatic. Recently, attention has turned to the gut-brain axis, a complex network of communication between the gastrointestinal tract and the brain, as a potential player in the pathogenesis of AD.
Here we exploited X-ray phase contrast tomography to provide an in-depth analysis of the link between the gut condition and AD, exploring gut anatomy and structure in murine models.
We conducted a comprehensive analysis by comparing the outcomes in various mouse models of cognitive impairment, including AD, frail mice, and frontotemporal dementia (FTD) affected mice. We discovered an association between substantial changes in the gut structure and the presence of amyloid-beta (Aβ) in the brain.  In particular, we investigated the gut morphology, the alterations of gut permeability and neurons distribution in the ileum.
Understanding the intricate interplay between gut condition and dementia may open new avenues for an early AD diagnosis and treatment offering hope for a future where these diseases may be more effectively treated.

**INTRODUCTION**
Alzheimer's disease (AD), the most common form of dementia, is a progressive neurodegenerative disorder characterized by synaptic dysfunction, cognitive decline, and a series of brain alterations including synaptic loss, chronic neuroinflammation and neuronal cell death (REF). As the prevalence of this devastating condition continues to rise, unraveling the underlying mechanisms and identifying novel therapeutic targets become paramount.
In recent years, there has been growing evidence to support a bidirectional communication between the gut and the brain [1-5], involving various interconnected pathways, such as neuroendocrine, immune, and neuronal. Dysfunction in this axis has been implicated in several psychiatric and neurological disorders including AD (REF).

The gut microbiota, which refers to the vast community of microorganisms residing in the intestinal tract, has been found to have profound influences on brain function and behavior.
Scientists have discovered that changes in the gut microbiota composition (namely dysbiosis) can contribute to AD development and progression [1-5].  Theoretically, dysbiosis foster the production of toxic metabolites, promoting inflammation and, consequently, the breakage of the gut/brain barriers. Bad bacteria could contribute to the disease by entering the circulation, reaching the brain, and initiating the classical AD-related cascade of neuropathological events. Moreover, altered gut microbiota produce various neurotransmitters and neuroactive substances that negatively influence brain function and behavior. These molecules can modulate synaptic plasticity, neuroinflammation, and amyloid-beta accumulation, which are all key features of AD.

Understanding the intricate crosstalk between the gut and the brain in AD has opened new possibilities for therapeutic interventions. Researchers are exploring the potential of modulating the gut microbiota through dietary interventions, probiotics, prebiotics, and fecal microbiota transplantation to improve cognitive function and alleviate the pathological processes associated with AD (REF).

This paper presents a groundbreaking application of X-ray phase-contrast tomography (XPCT) [6-9] to investigate the intricate relationship between AD and gut alterations. XPCT allowed us to visualize the entire intestinal structures and their microarchitecture, cells, and processes organization in 3D, providing a comprehensive view of the alterations associated with AD. We performed micro-XPCT at ANATOMIX at Soleil synchrotron and holo-nano-XPCT at ID16a at ESRF synchrotron.

For the first time, high-resolution 3D images of alteration in the intestine has been analyzed in detail, providing unprecedented insights into the gut-brain axis and its role in disease progression.

**RESULTS**:

*Morphology of villi*
To deeply investigate the morphological changes in the gut morphology associated with neurodegenerative disease, exploiting XPCT capabilities, we effectively measured the ileum villi of various mouse models. Specifically, we examined the distribution and morphology of villi in the control healthy groups C57BL/6 or SAMR1 mice, the latter as a normal aging control, compared to four specific models: APP/PS1, APP23, SAMP8 and P301S.
The following table shows the number of samples analyzed for the different groups:

| Groups      | Age       | Model    | Mouse number |
|-------------|-----------|----------|--------------|
| AD-Familiar | 18 months | APP/PS1  | 3            |
|             |           | APP23    | 3            |
|             |           | C57BL/6  | 3            |
|             |           |          |              |
|             |           |          |              |
| AD-Sporadic | 11 months | SAMP8    | 8            |
|             |           | SAMR1    | 5            |
|             |           | C57BL/6  | 7            |

*Table 1: Samples analyzed in this work.*

Both APP/PS1[10] and APP23 [11] mice are animal models used to study AD, carrying a typical AD human mutation, and developing AD phenotype including cognitive impairment and brain Aβ plaque deposition, neuronal damage, neuroinflammation, and brain atrophy. While SAMP8 [12] and SAMR1 [13] transgenic mice were used as models for cognitive impaired to fast ageing, with SAMP8 often being associated with sporadic AD, and SAMR1 as a control of SAMP8. P301S [14] is a genetically modified mouse model mimicking FTD disease (Fronto-Temporal Dementia), which develops cognitive deficits, neuronal alterations, neuroinflammation, and tauopathy.

Our simultaneous investigation on corresponding brains of mice whose gut were studied, revealed that Aβ plaques were only visible in APP/PS1 and APP23, as expected. Figures 1a illustrate two hemibrains: on the left an APP/PS1 brain where Aβ plaques are well visible in the cortex, while on the right a SAMP8 brain without plaques, similarly to the control, SAMR1 or P301S brains.

XPCT images of mouse guts are reported in Figure1b, clearly displaying notable differences in shape and density between APP/PS1or APP23, and the other mouse models.

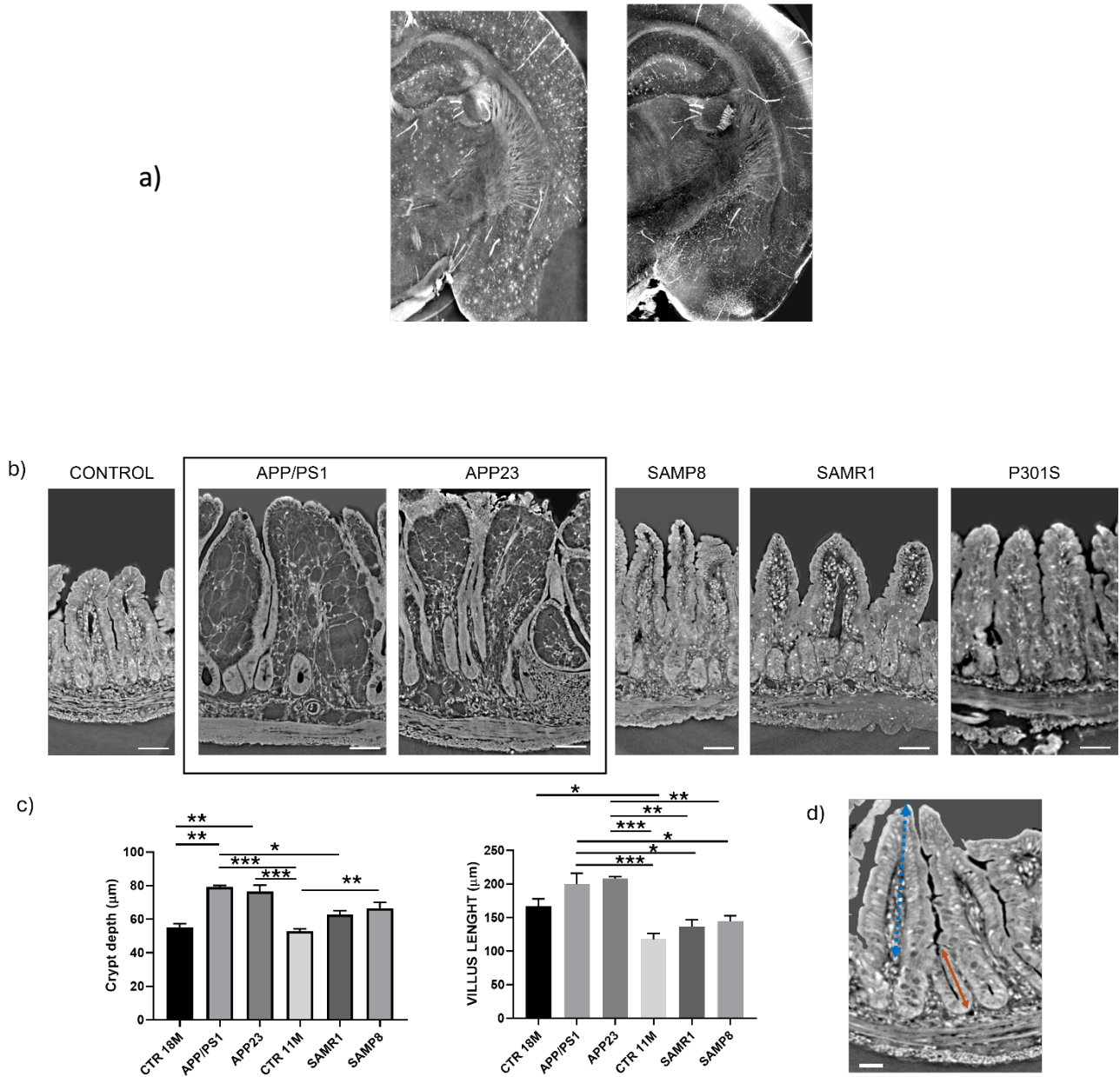

*Figure 1: Experiment carried out at ANATOMIX at Soleil.*
*a) X-ray Phase Contrast Tomography (XPCT) of hemibrains, on the right an example of APP/PS1 model while on the left a control one. b) Morphology of the ileum villi in different dementia models and their control groups. The scale bar is 50 μm. c) Quantification of crypt depth (left side) and villus length (right side) in the different mouse models. Results are shown as mean ± SEM. One way ANOVA p<0.0001; Post hoc by Tukey's HSD test: *p<0.05; **p<0.005; *** p<0.0001. d) Scheme of villus length (blue dotted arrow) and crypt depth (orange arrow) measurements. Scale bar = 20um.*

Crypt depth and villus length were determined taking into account the boundary between the villi and the crypts. The criterion described in the literature [*Juha Taavela, Keijo Viiri: "Apolipoprotein A4 Defines the Villus-Crypt Border in Duodenal Specimens for Celiac Disease Morphometry"*] was therefore adopted to precisely define the boundary between these two region. To improve the reliability of measurements, two observers (authors F.P. and N.M.) quantified the parameters independently. To avoid measurement errors, the length of the villi was measured at the point of maximum elongation, as it was possible to work on the virtual volume.

Morphometric quantification of the crypt depth and villus length revealed significant differences among mice (Figure 1c). In particular, both APP/APS1 and APP3 mice showed significantly deeper crypt if compared to control mice at the same age. Moreover, also SAMP8 mice exhibited significantly deeper crypt when compared to their age-matching controls, confirming an effect of AD on the shape of murine crypts. As for the villus length, significant differences emerged between different transgenic mouse models and between controls at different ages, in which older controls showed longer villus length than young controls. These data imply an effect of AD disease also on this parameter, although no significant differences were detected between the transgenic models and the age-matching controls.

The thickness of the epithelium barrier in the AD models was considerably reduced, while the villus region of lamina propria appeared larger. Notably, the lamina propria in APP/PS1 and APP23 (whose corresponding brains show Aβ plaques) exhibited lower density – darker shades of the greyscale - compared to both the control group and the other models. The epithelium displayed a more corrugated shape, suggesting that it could be more susceptible to attacks from harmful micro-bacteria within the microbiota. Notably, the lamina propria in APP/PS1 and APP23 (whose corresponding brains show Aβ plaques) exhibited lower density – darker shades of the greyscale - compared to both the control group and the other models. The epithelium displayed a more corrugated shape, suggesting that it could be more susceptible to attacks from harmful micro-bacteria within the microbiota. The histological analysis performed by hematoxylin and eosin staining validate the features observed in XPCT images (Figure 2). Indeed, ileum of control healthy mice showed well-organized villi, characterized by the surface of columnar epithelial enterocytes with a striated border, needed for optimal nutrient absorption, and the lamina propria, a highly cellular loose connective tissue that makes up the core of the villus, containing mostly lymphocytes (Figure 2d, e, f). APP23 ileum, instead, appears to have a flat luminal surface of the villi, with a cuboidal striated epithelium and a lamina propria completely disorganized, containing round-shape subcellular structures (Figure 2l, m, n).

Histological features of APP23 ileus in part resemble that one described in the colon of Tg2576 mice, another transgenic murine model used to study AD (PLoS ONE 14(4): e0215205. https://doi.org/10.1371/journal.pone.0215205)

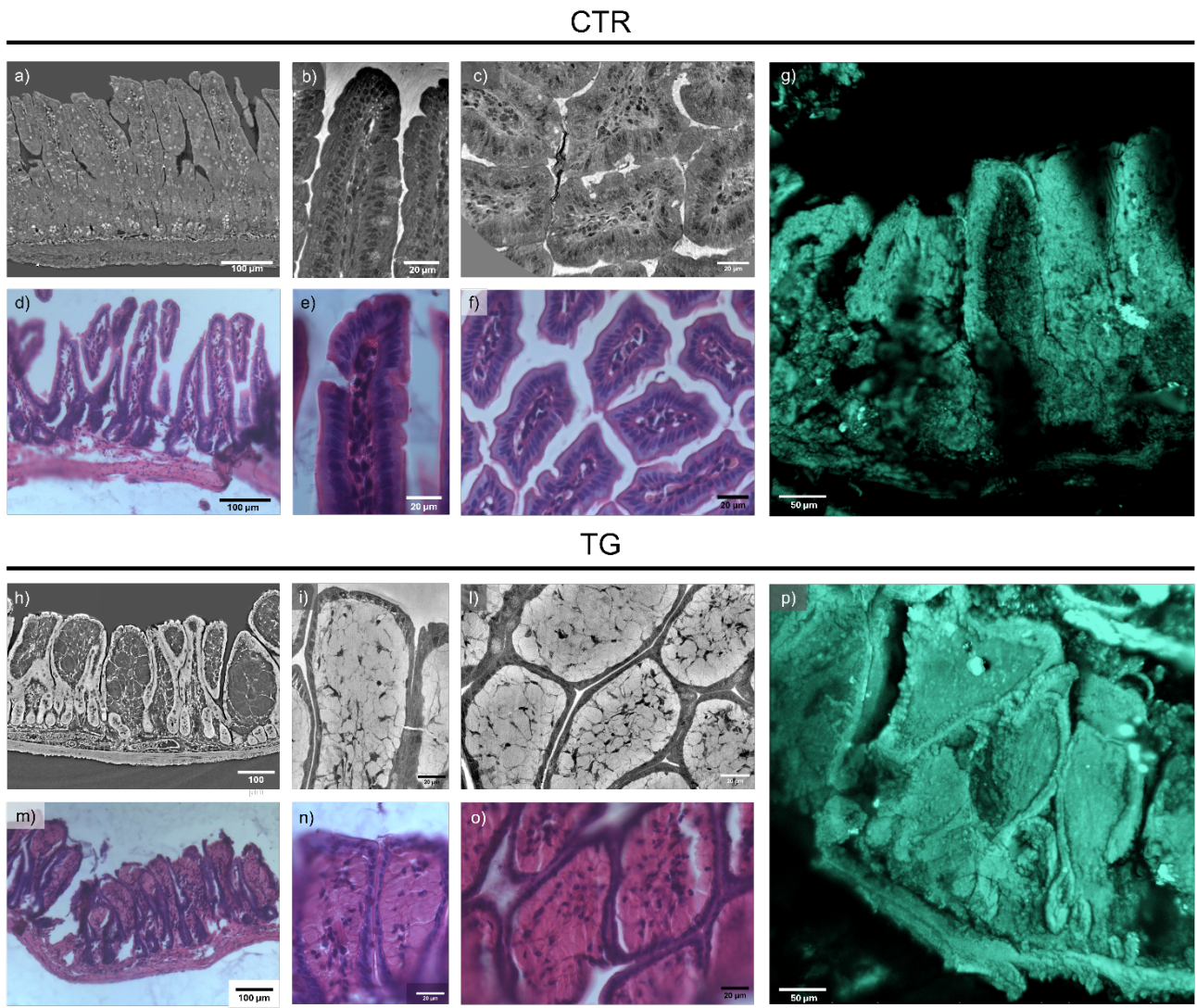

*Figure 2: Comparison between XPCT and histology. XPCT images of control ileum (a-c) and of APP23 ileum (h-l), compared to histological sections stained with hematoxylin and eosin (d-f) (m-o) and to confocal microscopy images (g) (h). a and g were acquired with micro-XPCT; b, c, h, i were acquired with holo-nano-tomography.*

To deeper analyze the cellular component, we conducted holo-nano-tomography (HNT) of the ileum with a pixel size of 80 nm of the APP/PS1, SAMP8, and SAMR1 mouse guts, as well as of the C57BL/6 control ones. Experiment was carried out at ID16A at ESRF synchrotron. Representative findings obtained from a APP/PS1 sample has been shown in Figure 3, which represents a lateral view of the ileum structure at the level of the crypts and the tunica muscolaris. Different tissues and cellular types were highlighted in circles.

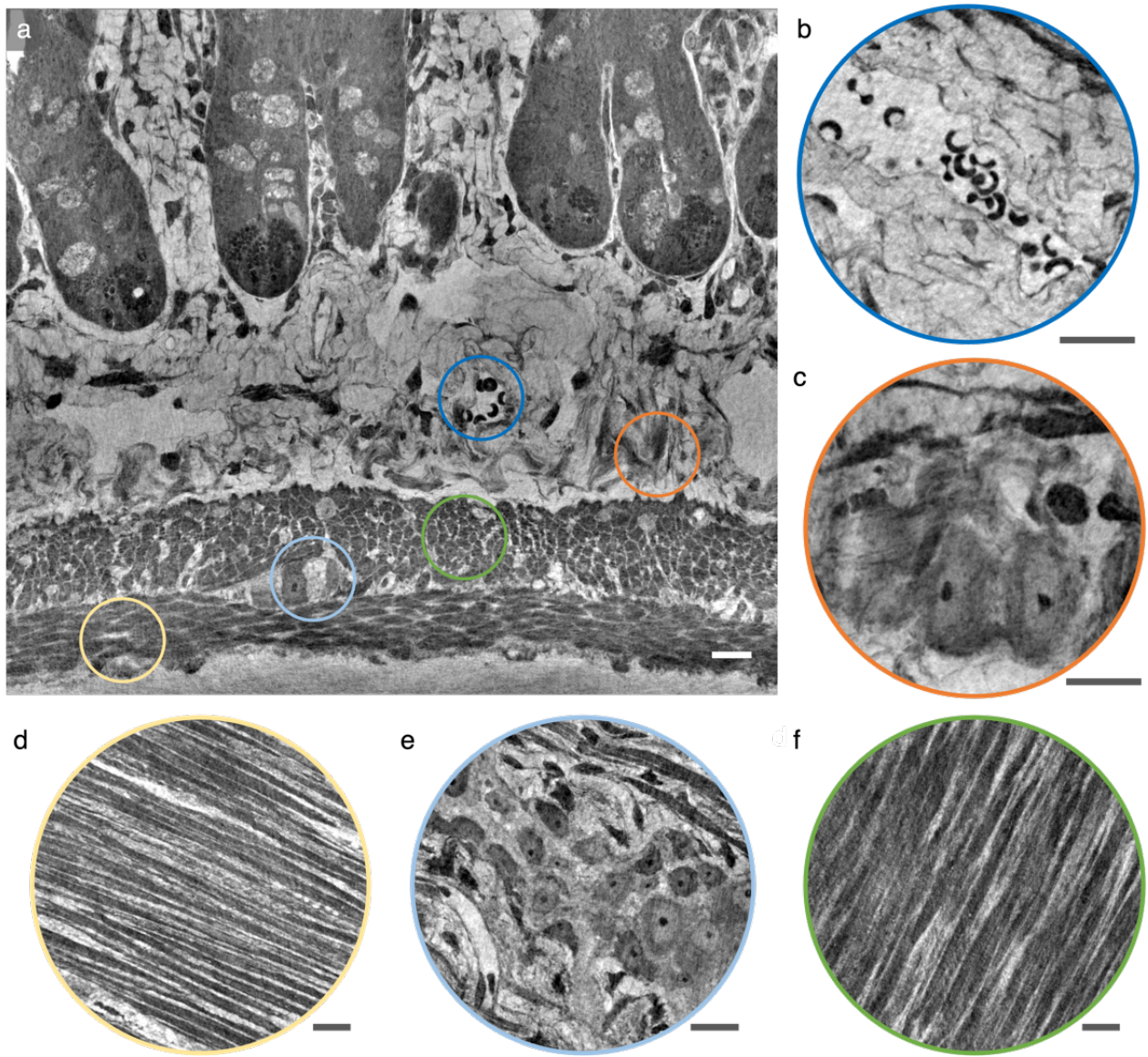

*Figure 3: Holo-nano-tomography of a APP/PS1 mouse ileum. Lateral view of the APP/PS1 ileum structure at the level of the criptae and the tunica muscolaris. The circles indicate the approximate position of the morphological details shown in the figure. b) Blood vessels with blood cells. c) Neurons belonging to the submucosal plexus (Meissner). d) Longitudinal layer of the tunica muscolaris. e) Neurons of the myenteric plexus (Auerbach). f) Circular layer of the tunica muscularis. Scale bar: 10 µm. Experiment carried out at ID16A at ESRF.*

*Secretory epithelial cells*
We also investigated the secretory epithelial cells part of the innate immune system, namely the Paneth and goblet cells. Specifically, Paneth cells play a role in innate immunity by producing antimicrobial substances, while goblet cells contribute to intestinal protection through the secretion of a different type of mucus that allows the passage of certain nutrients. When the function of these cells is compromised, it leads to the appearance of unprotected regions of the gut epithelium (REFERENZA).
To investigate their possible role in gut alteration in disease, we compared Paneth cell abundance in the different models. This quantitative analysis was carried out in 3D, which is possible only with XPCT technique. The 3D image of the crypt, presented in Figure 4a, showcase the remarkable visibility of Paneth and goblet cells. Quantitative analyses showed a significant difference in the number of the Goblet cells per crypt among the different mouse models (Figure 4b), suggesting an effect of the disease. While no significant differences were detected in the number of Paneth cells (Figure 4c). Moreover, while APP/PS1 mice show an increase in Goblet cells and a decrease of the Paneth ones, in the SAMP8 mouse model these findings are opposite.

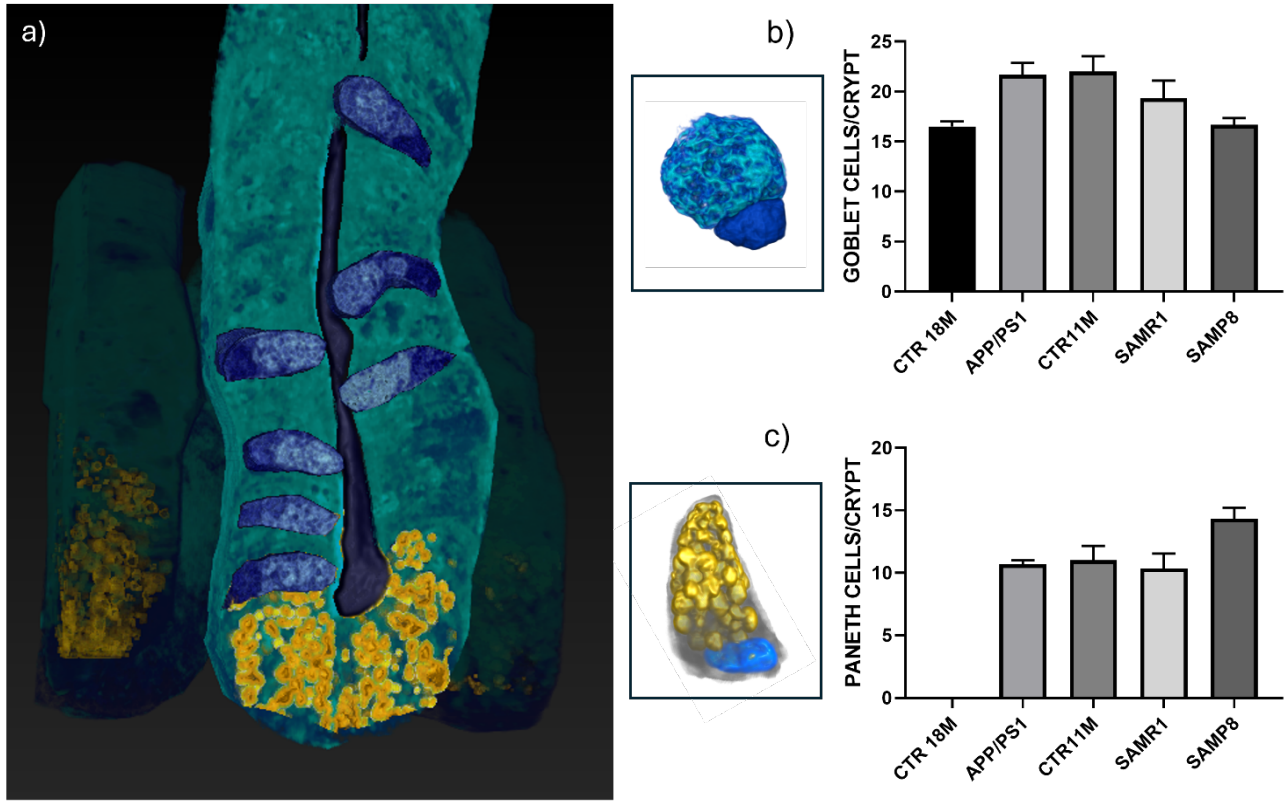

*Figure 4: XPCT analysis of the Paneth and goblet cells. a) Representative 3D rendering of one crypt of SAMP8 mouse. In yellow are indicated the Paneth cells and in blue the goblet ones. Quantitative analysis of goblet (b) and Paneth (c) cells in different AD and control models. Results are shown as mean ± SEM. One way ANOVA p<0.05 for b) and c).*

*Lymphatic follicle hypertrophy*
XPCT technique was employed also to study the Peyer's patches. We found a hyperplasia of the Peyer's patches, indicative of an inflammatory state which may reduce the permeability of the gut barrier. Figure 5a shows XPCT images of Peyer patches in hyperplasia (left) and normal condition (right). Figure 5b shows the percentage of mice presenting hyperplasia (dark green) over the total number of mice analyzed for each group. We observed an higher percentage of the enlarged patches in the SAMP8 samples than in the other groups.

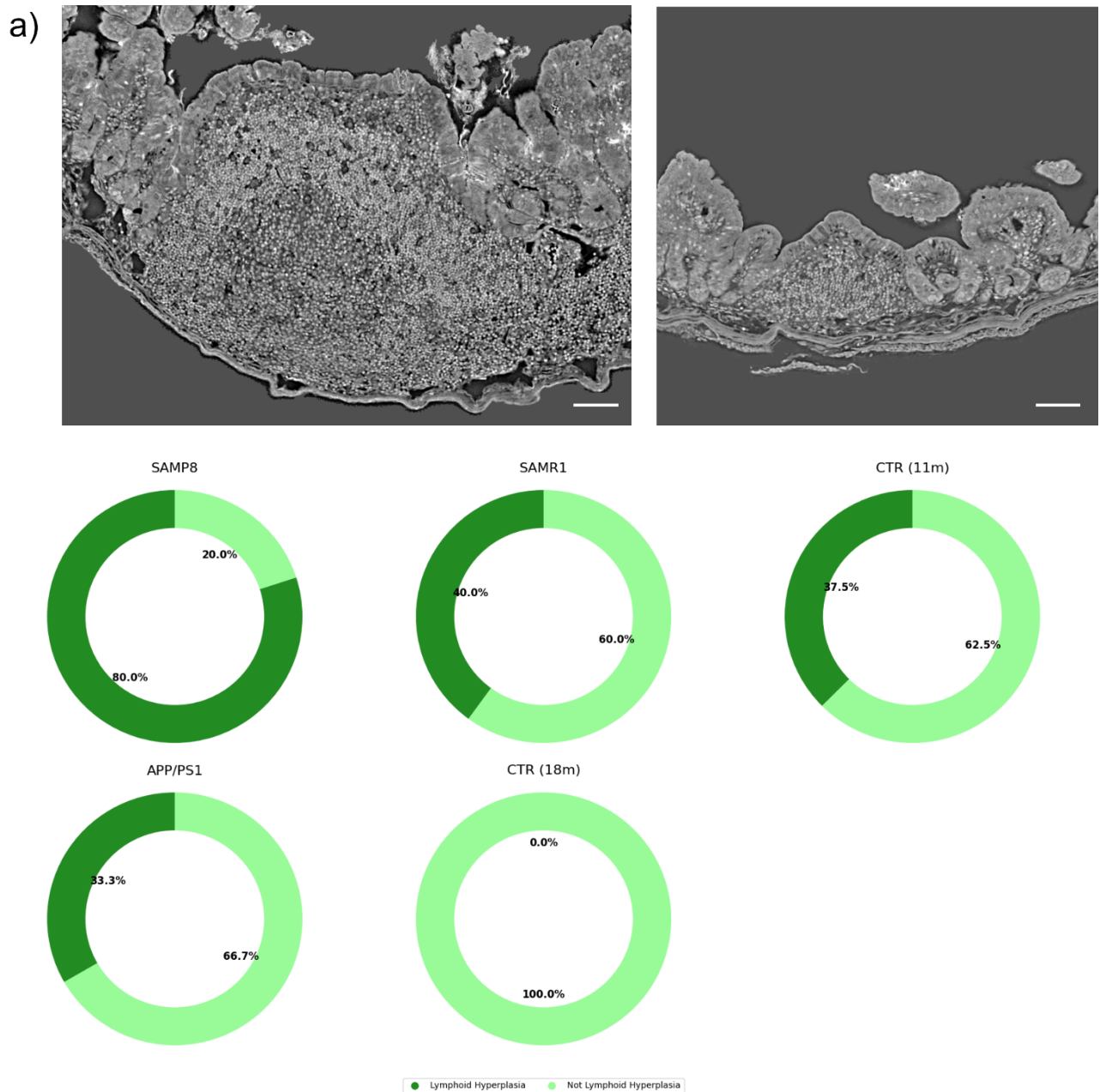

*Figure 5: XPCT analysis of Peyer's patches. a) XPCT images of Peyer's patches in hyperplasia (left) and the control condition (right). Scale bar = 50 um. Experiment carried out at ANATOMIX at Soleil. b) Percentage of hyperplastic samples for each group of dementia and control mice.*

*Telocytes*
On the basis of the morphology and position, it is possible to distinguish the telocytes [10-11], a type of interstitial (stromal) cells. Figure 6 is an impressive image of the telocytes in the APPS1 model. Telocytes (in green) present very long and very thin prolongations, the telopodes, indicated in red, which can be seen very well on XPCT, but are mostly below the resolving power of light microscopy. They are mainly involved in the differentiation and proliferation of intestinal crypt stem cells and interact with other cell types (macrophages, smooth muscle cells, other immune cells) for various purposes, e.g, mediating chemotaxis of immune cells. The immune cells are colored in blue.

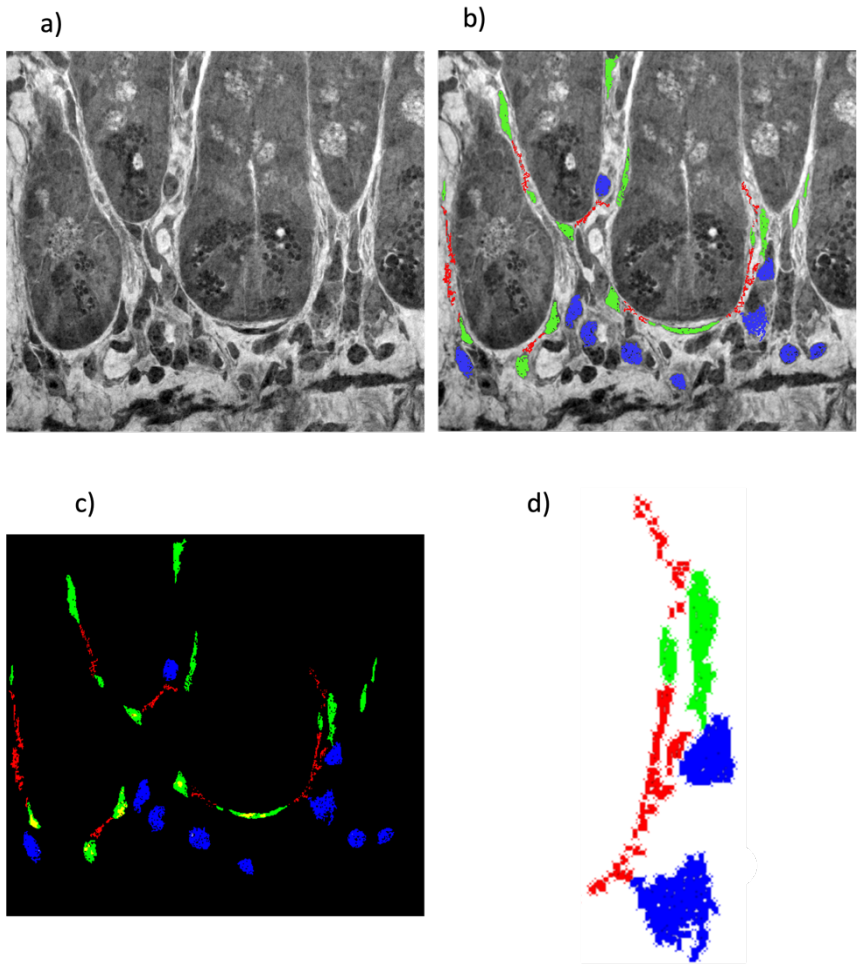

*Figure 6: XPCT analysis of telocytes. a) Reperesentative longitudinal view of the crypts and villi of the APPS1 model. b) and c) Telocytes are colored in green and the long process (telopodies) are indicated in red; immune cells in blu. d) Telocytes, telopods and immune cells interaction. Experiment carried out at ID16A at ESRF.*

Figure 7 shows the 3D view of the telocytes forming a 3D network surrounding the crypts. The 3D image reveals that the shape of the telocyte cells is flat and wide in APP/PS1 ileum, and it is only their positions relative to the crypt that make them appear narrow and elongated.

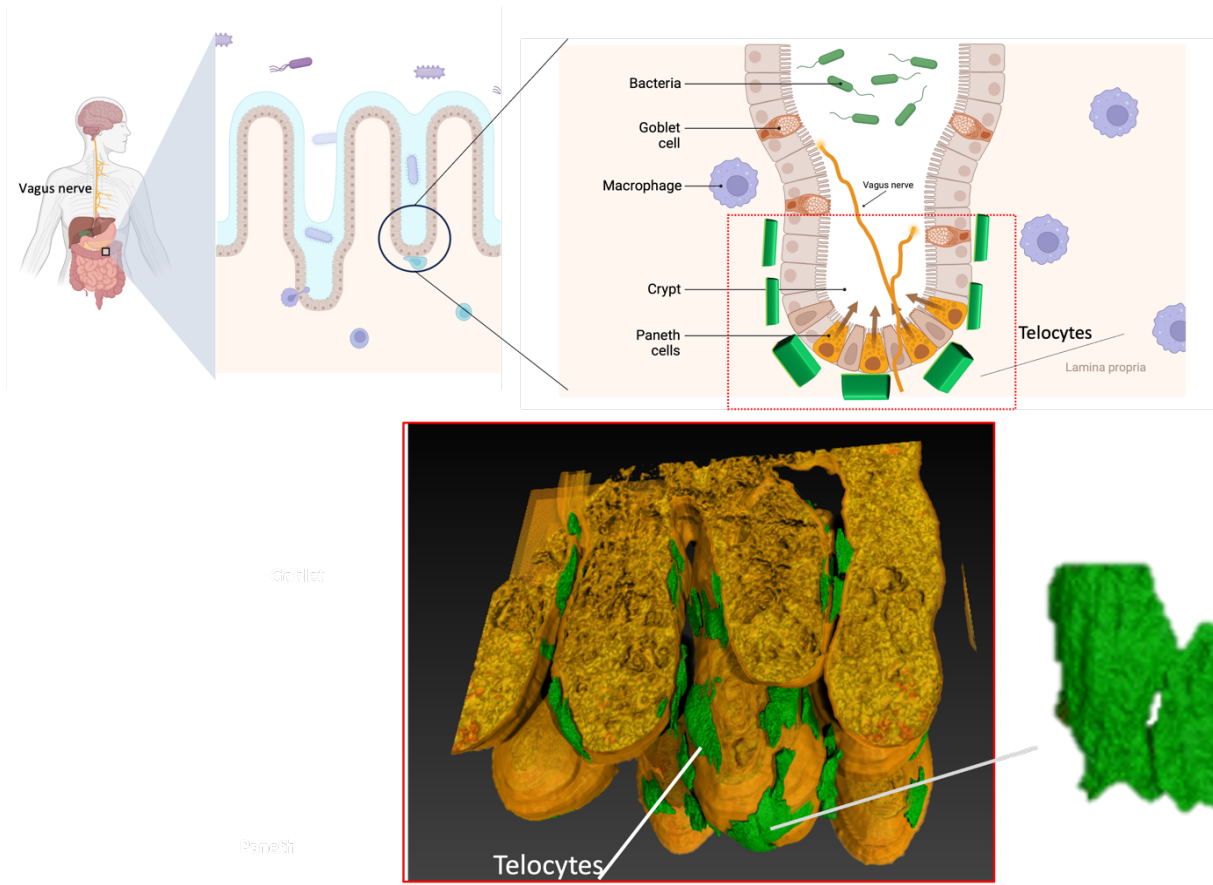

*Figure 7: XPCT analysis of telocyte shape. a) …*

## METHODS

*Animals.*
**APP/PS1.** APPswe/PS1dE9 transgenic male mice [B6C3 – Tg(APPswe, PSEN1dE9)85Dbo/Mmjax mice], which express a chimeric mouse/human amyloid precursor protein (Mo/HuAPP695swe) and a mutant human presenilin 1 (PS1-dE9), were purchased from Jackson Laboratories (United States).

APP23

SAMP8

SAMR1

P301S

All animals were housed in a SPF facility in groups of 4 in standard mouse cages containing sawdust with food (2018S Envigo diet) and water ad libitum, under conventional laboratory conditions (room temperature: 20 ± 2°C; humidity: 60%) and a 12/12 h light/dark cycle. The Institute of Pharmacological Research Mario Negri adheres to the principles set out in the following laws, regulation, and policies governing the Care and Use of Laboratory Animals: Italian Governing Law

(D.lgs 26/2014; Authorization n. 19/2008-A issued March 6, 2008 by Ministry of Health); Mario Negri Institutional Regulations and Policies providing internal authorization for persons conducting animal experiments (Quality Management System Certificate – UNI EN ISO 9001:2015 – Reg. N° 6121); the NIH Guide for the Care and Use of Laboratory Animals (2011 edition) and EU directives and guidelines (EEC Council Directive 2010/63/UE). The statement of Compliance (Assurance) with the Public Health Service (PHS) Policy on Human Care and Use of Laboratory Animals has been reviewed (9/9/2014; Animal Welfare Assurance #A5023-01). The mice were used for the experiment at 18 months of age. Mice were killed under $CO_2$ inhalation.

*Sample preparation*.

Samples were dehydrated through a graded ethanol series (70/95/100%), put in propylene oxide, and included in paraffin.

*Micro XPCT.*

The XPCT experiments were performed (1) at the beamline ID19 of the European Synchrotron Radiation Facility (ESRF, Grenoble, France), (2) at the ANATOMIX beamline of Synchrotron SOLEIL (Paris, France) in free-space propagation mode.

(1) Data acquisition at ID19 was carried out using pink beam with an energy peaked around 26.5 keV. The sample-detector distance was set at 10 cm. The detector had an effective pixel size of 0.65 μm. The tomography was produced by means of 3600 in extended field-of-view (FOV) mode, an experimental procedure of acquisition which allows to almost double the effective horizontal width of the FOV of the detector. The rotation axis is moved close to either left or right side of the FOV and a dataset of projections, having size equal to the detector FOV, is collected over 360°. After properly stitching the sinograms, the reconstruction procedure can be performed as usual. The acquisition time for each angular position was 150 ms. Data preprocessing, phase retrieval and tomographic reconstruction were performed with Tomwer software provided by the ESRF.

(2) The XPCT experiment at ANATOMIX beamline was performed with a filtered white beam peaked around 20 keV. The measurements were performed with an effective pixel size of 3.08 μm and 0.65 μm, resulting from 2× and 10×optics coupled with Orca Flash 4.0 camera (sensor type CMOS, sensor array size 2048 × 2048, pixel size 6.5 micron 16-bit nominal dynamic range). The propagation distance was set to 1 m and at 20 mm respectively. The experiment was carried out recording 4000 projections in extended field-of-view (FOV) mode covering a total angle range of 360°, The exposure time was 150 ms and 250 ms per projection, with 2× and 10× optics respectively. Data pre-processing, phase retrieval and tomographic reconstruction were performed with PyHST software package.Phase retrieval was performed by using Paganin's algorithm, since we acquired a single defocused image per angle in free-space propagation, and radiation wavelength and material density met the requirements. This method allows the simultaneous extraction of phase and amplitude of the wave as the ratio of the phase term over the absorption term. Tomographic reconstruction of the 3D volumes was performed from the phase- retrieved angular projections using the Filtered Back Projection algorithm.

In the gray scale of the tomographic images, white corresponds to higher density structures and black to lower density structures.

### The holographic (holo-)nano-XPCT.

The holographic (holo-)nano-XPCT experiment was carried out at Nano-Imaging ID16A beamline of the ESRF. A pair of multilayer-coated Kirkpatrick-Baez optics was used to focus the X-rays (~30 nm) at 17 keV. The sample is put in the divergent beam downstream of the focus to produce magnified phase contrast images. The projection geometry also allows zooming into specific regions of a large sample by combining scans with different magnifications and FOV ([Mokso et al., 2007](); [Bartels et al., 2015]()). By measuring the Fresnel diffraction patterns at different effective propagation distances, the phase maps of the sample can be retrieved via holographic reconstruction, this so-called phase-retrieval procedure ([Cloetens et al., 1999]()) being implemented using GNU Octave software. Magnified radiographs were recorded onto an X-ray detector using a FReLoN-charged coupled device. For one tomography scan, 1500 projections were acquired with 0.32 s exposure time and 50 nm effective pixel size. Tomography scans at four different foci-to-sample distances were acquired to complete one holotomography scan. The tomographic reconstruction was obtained with ESRF PyHST software package. In this kind of images, the shades of gray are proportional to electron density, with black corresponding to the highest value of the density spectrum, whereas white corresponds to the lowest value, hence to features of lowest density.

### Image analysis.

Image analysis and segmentation were performed using ImageJ. 3D rendering images were obtained using the high-end software VGstudio Max. Before extracting qualitative and quantitative information, images were pre- preprocessed to eliminate or mitigate artifacts due to experimental conditions or computational reconstruction. Ring artifacts were removed by an improved frequency filtering.(REF).

To enhance the contrast and visualize structures developing in 3D and therefore lying on different tomographic slices, we exploited the z-projection of maximum intensities, which consists in projecting on the visualization plane the voxels of a set of continuous slices. Each pixel of the output image contains the maximum value found along the axis perpendicular to that pixel.

### Histological analyses.
Ileum fragments from C57BL/6 and APP3, included in paraffin, were cut by using a Leica microtome (Leica, Wetzlar, Germany) in 10 µm-thick sections. Hematoxylin and eosin (Sigma-Aldrich, Saint Louis, MO, USA) staining was performed according to the standard manufacturer's instructions. Briefly, after deparaffinization to water, slides were stained for 10 minutes with Mayer's Hematoxylin Solution. Slides were then washed under tap water for 10 minutes and counterstained with Eosin Y Solution for few seconds. After removing the excess of Eosin, sections were dehydrated, and cover slips were mounted with Eukitt Mounting Medium.
Representative pictures were acquired by using an Axioplan microscope (Zeiss, Germany) with Axiocam 503 color and ZEN 3.3 (blue edition) software.

### Statistics.
The differences between mouse models were analyzed in independent litters, by considering as "n" an independent biological sample, not a technical replicate. After verifying the normal distribution of the data and their homoscedasticity by means of Levene's test, data were analyzed with ONE-way ANOVA, followed by Tukey's HSD as post hoc test. All values were expressed as mean ± standard error of the mean (SEM).

# REFERENCES

1) J Alzheimers Dis 2017;58(1):1-15. doi: 10.3233/JAD-161141. The Gut Microbiota and Alzheimer's Disease

2) *Molecular Neurodegeneration* volume 18, Article number: 9 (2023). The gut microbiome in Alzheimer's disease: what we know and what remains to be explored

3) *Scientific Reports* volume 13, Article number: 5258 (2023). Genetic correlations between Alzheimer's disease and gut microbiome genera

4) *Brain*, Volume 146, Issue 12, December 2023, Pages 4916–4934, *Microbiota from Alzheimer's patients induce deficits in cognition and hippocampal neurogenesis*

5) *The gut microbiome in Alzheimer's disease: what we know and what remains to be explored.* Chandra, S., Sisodia, S.S. & Vassar, R.J. *Mol Neurodegeneration* **18**, 9 (2023).

6) Massimi, L.; Bukreeva, I.; Santamaria, G.; Fratini, M.; Corbelli, A.; Brun, F.; Fumagalli, S.; Maugeri, L.; Pacureanu, A.; Cloetens, P.; et al. Exploring Alzheimer's disease mouse brain through X-ray phase contrast tomography: From the cell to the organ. *NeuroImage* **2018**, *184*, 490–495.

7) Di Fonzo, S.; Jark, W.; Soullié, G.; Cedola, A.; Lagomarsino, S.; Cloetens, P.; Riekel, C. Submicrometre resolution phase-contrast radiography with the beam from an X-ray waveguide. *J. Synchrotron Radiat.* **1998**, *5 Pt 3*, 376–378.

8) Mittone, A.; Fardin, L.; Di Lillo, F.; Fratini, M.; Requardt, H.; Mauro, A.; Homs-Regojo, R.A.; Douissard, P.A.; Barbone, G.E.; Stroebel, J.; et al. Multiscale pink-beam microCT imaging at the ESRF-ID17 biomedical beamline. *J. Synchrotron Radiat.* **2020**, *27 Pt 5*, 1347–1357.

9) A Cedola, G Campi, D Pelliccia, I Bukreeva, M Fratini, M Burghammer, ...
Three dimensional visualization of engineered bone and soft tissue by combined x-ray micro-diffraction and phase contrast tomography. Physics in Medicine & Biology **2013** 59 (1), 189.

10) Waldron AM, Fissers J, Van Eetveldt A, Van Broeck B, Mercken M, Pemberton DJ, Van Der Veken P, Augustyns K, Joossens J, Stroobants S, Dedeurwaerdere S, Wyffels L, Staelens S. In Vivo Amyloid-β Imaging in the APPPS1-21 Transgenic Mouse Model with a (89)Zr-Labeled Monoclonal Antibody. Front Aging Neurosci. 2016 Mar 31;8:67. doi: 10.3389/fnagi.2016.00067. PMID: 27064204; PMCID: PMC4815004.

11) Van Dam D, Vloeberghs E, Abramowski D, Staufenbiel M, De Deyn PP. APP23 mice as a model of Alzheimer's disease: an example of a transgenic approach to modeling a CNS disorder. CNS Spectr. 2005 Mar;10(3):207-22. doi: 10.1017/s1092852900010051. PMID: 15744222.

12) Liu B, Liu J, Shi JS. SAMP8 Mice as a Model of Age-Related Cognition Decline with Underlying Mechanisms in Alzheimer's Disease. J Alzheimers Dis. 2020;75(2):385-395. doi: 10.3233/JAD-200063. PMID: 32310176.

13) Shimada A, Hasegawa-Ishii S. Senescence-accelerated Mice (SAMs) as a Model for Brain Aging and Immunosenescence. Aging Dis. 2011 Oct;2(5):414-35. Epub 2011 Oct 28. PMID: 22396891; PMCID: PMC3295080.

14) Takeuchi H, Iba M, Inoue H, Higuchi M, Takao K, Tsukita K, Karatsu Y, Iwamoto Y, Miyakawa T, Suhara T, Trojanowski JQ, Lee VM, Takahashi R. P301S mutant human tau transgenic mice manifest early symptoms of human tauopathies with dementia and altered sensorimotor gating. PLoS One. 2011;6(6):e21050. doi: 10.1371/journal.pone.0021050. Epub 2011 Jun 15. PMID: 21698260; PMCID: PMC3115982.

15) Lucio Díaz-Flores et al Telocytes in the Normal and Pathological Peripheral Nervous System. Int. J. Mol. Sci. 2020, 21, 4320; doi:10.3390/ijms21124320

16) Nicolae Mirancea et al. Rom J Morphol Embryol 2022, 63(2):335–347

## Acknowledgements

We acknowledge the European Synchrotron Radiation Facility for provision of synchrotron radiation facilities, and we would like to thank the staff for assistance in using beamline ID 16a and for constructive discussion. ANATOMIX is an Equipment of Excellence (EQUIPEX) funded by the Investments for the Future program of the French National Research Agency (ANR), project NanoimagesX, grant no. ANR-11-EQPX-0031. (proposal 20190809).

This work was supported by Regione Puglia and CNR for Tecnopolo per la Medicina di Precisione. D. G. R. n. 2117 of 21.11.2018. It was also supported by *PIANO NAZIONALE DI RIPRESA E RESILIENZA (PNRR) - missione 4 componente 2 investimento1.5, progetto ecs 00000024 Rome Technopole*